\RequirePackage{filecontents}

\RequirePackage{snapshot}
\documentclass[aps,prl,reprint,secnumarabic]{preprint}

\renewcommand\documentclass[2][]{}
\let\origparagraph=\paragraph
\AtBeginDocument{%
  \let\paragraph=\origparagraph
}
\makeatother
\documentclass[reprint,aps,prl,floatfix,longbibliography]{revtex4-1}

\usepackage[T1]{fontenc}
\usepackage{amsmath}
\usepackage{textcomp}
\usepackage{txfonts}
\usepackage{tgtermes}
\usepackage[altg]{qtxmath}
\renewcommand*{\vec}[1]{\boldsymbol{#1}}
\usepackage{microtype}
\usepackage{graphicx}
\usepackage{braket}
\usepackage[usenames,dvipsnames]{color}
\graphicspath{{figures/}}
\usepackage[breaklinks=true,
pdfauthor={Meng Wen, Heiko Bauke, and Christoph H. Keitel},
pdftitle={Identifying the Stern-Gerlach force of classical electron dynamics}]{hyperref}
\setcitestyle{super,sort&compress}

\newcommand*{\D}{\mathrm{d}}
\newcommand*{\e}{\mathrm{e}}
\newcommand*{\I}{\mathrm{i}}
\newcommand*{\trans}[1]{#1^{\mathsf{T}}}

\makeatletter
\AtBeginDocument{\renewcommand{\@biblabel}[1]{#1.}}
\def\@bibdataout@aps{%
 \immediate\write\@bibdataout{%
  @CONTROL{%
   apsrev41Control%
   \longbibliography@sw{%
    ,author="0a",editor="1",pages="1",title="0",year="0"%
   }{%
    ,author="0a",editor="1",pages="0",title="",year="1"%
   }%
  }%
 }%
 \if@filesw
  \immediate\write\@auxout{\string\citation{apsrev41Control}}%
 \fi
}%
\makeatother

\begin{document}

\title{Identifying the Stern-Gerlach force of classical electron dynamics}

\author{Meng Wen}
\affiliation{Max-Planck-Institut f{\"u}r Kernphysik, Saupfercheckweg~1, 69117~Heidelberg, Germany}
\author{Heiko Bauke}
\email{heiko.bauke@mpi-hd.mpg.de}
\affiliation{Max-Planck-Institut f{\"u}r Kernphysik, Saupfercheckweg~1, 69117~Heidelberg, Germany}
\author{Christoph H. Keitel}
\affiliation{Max-Planck-Institut f{\"u}r Kernphysik, Saupfercheckweg~1, 69117~Heidelberg, Germany}

\begin{abstract}
  Different classical theories are commonly applied in various branches
  of physics to describe the relativistic dynamics of electrons by
  coupled equations for the orbital motion and spin precession.
  Exemplarily, we benchmark the Frenkel model
  and the classical Foldy-Wouthuysen model
  with spin-dependent forces (Stern-Gerlach forces) to the quantum
  dynamics as predicted by the Dirac equation.  Both classical theories
  can lead to different or even contradicting predictions how the
  Stern-Gerlach forces modify the electron's orbital motion, when the
  electron moves in strong electromagnetic field configurations of
  emerging high-intensity laser facilities.  In this way, one may
  evaluate the validity and identify the limits of these classical
  theories via a comparison with possible experiments to provide a
  proper description of spin-induced dynamics.  Our results indicate
  that the Foldy-Wouthuysen model is qualitatively in better agreement
  with the Dirac theory than the widely used Frenkel model.
\end{abstract}


\date{August 22, 2016}

\maketitle

\allowdisplaybreaks

The electron couples to external electromagnetic fields via its charge
as well as via its spin.  Gradients of the electromagnetic fields induce
a spin-dependent force in addition to the Lorentz force.  Spin-dependent
motion is implemented in the seminal Stern-Gerlach experiment
\cite{Gerlach:Stern:1922:experimentelle_Nachweis} and variants thereof
\cite{Batelaan:Gay:etal:1997:Stern-Gerlach_Effect,McGregor:Bach:etal:2011:Transverse_quantum}.
Effects of spin-dependent forces appear in condensed matter
\cite{Shen:2005:Spin_Transverse}, in astrophysical systems
\cite{Mahajan:Asenjo:etal:2014:Comparison_of}, in quantum plasmas
\cite{Marklund:Brodin:2007:Dynamics_of,Brodin:Marklund:2007:Spin_magnetohydrodynamics},
and at relativistic electrons in strong electromagnetic fields
\cite{Hu:Keitel:1999:Spin_Signatures,Walser:Urbach:etal:2002:Spin_and,Roman:Roso:etal:2003:complete_description,Faisal:Bhattacharyya:2004:Spin_Asymmetry,Klaiber:Yakaboylu:etal:2014:Spin_dynamics,Zimmermann:Buller:etal:2015:Strong-Field_Excitation}.
A consistent theoretical framework for the description of particles with
internal angular momentum is provided by the Dirac equation
\cite{Dirac:1928:Quantum_Theory}. The application of this
quantum-mechanical theory, however, is not always feasible and/or
necessary if quantum effects are not important.  Classical models of
charged point-like particles with spin in electromagnetic fields are
appealing because they are usually simpler from a mathematical point of
view than the Dirac equation and are easier to interpret.  A first
covariant theory to describe the dynamics of a charged particle with
spin was proposed by Frenkel in 1926 by purely classical considerations
\cite{Frenkel:1926:Elektrodynamik_des}.  The Frenkel model has been
employed in many studies and continues stimulating new research
\cite{Holten:1991:On_electrodynamics,Yee:Bander:1993:Equations_of,Chaichian:Gonzalez-Felipe:etal:1997:Spinning_relativistic,Pomeranskii:Senkov:etal:2000:Spinning_relativistic,Gralla:Harte:etal:2009:Rigorous_derivation,Karabali:Nair:2014:Relativistic_Particle}.
Considering that the spin was introduced as an intrinsic quantum feature
of the electron \cite{Giulini:2008:spin} it may, however, appear
appropriate to start from quantum theory to find a classical model for
the electron.  Such a model can be derived from relativistic quantum
theory by applying the correspondence principle to the Heisenberg
equation for the time evolution of the position, the kinematic momentum,
and the spin in the Foldy-Wouthuysen representations of the Dirac
equation
\cite{Silenko:2008:Foldy-Wouthyusen_transformation,Chen:Chiou:2014:Correspondence_between}.
Both kinds of classical models are currently employed in different
branches of physics, e.\,g., 
the classical Foldy-Wouthuysen model in gravitational fields
\cite{Obukhov:Silenko:etal:2009:Spin_dynamics} or in crystals
\cite{Silenko:2015:Quantum-mechanical_description}, and the Frenkel
model in astrophysics \cite{Mahajan:Asenjo:etal:2014:Comparison_of} or
in plasma fields \cite{Flood:Burton:2015:Stern-Gerlach_surfing}.
Nevertheless, the validity and the limits of those classical models have
not been studied, e.\,g., by a comparison to the Dirac theory.  The lack
of a widely accepted classical description of the electron is deeply
related to interpretation problems regarding kinematic momentum
operators in the Dirac theory
\cite{Gurtler:Hestenes:1975:Consistency_in} and identifying an accurate
classical model may also facilitate insights into quantum theory and
will be valuable for systems where quantum descriptions are too complex
such as for many-particle systems in extreme laser pulses.

\section*{Results}

\begin{figure*}
  \centering
  \includegraphics[scale=0.475]{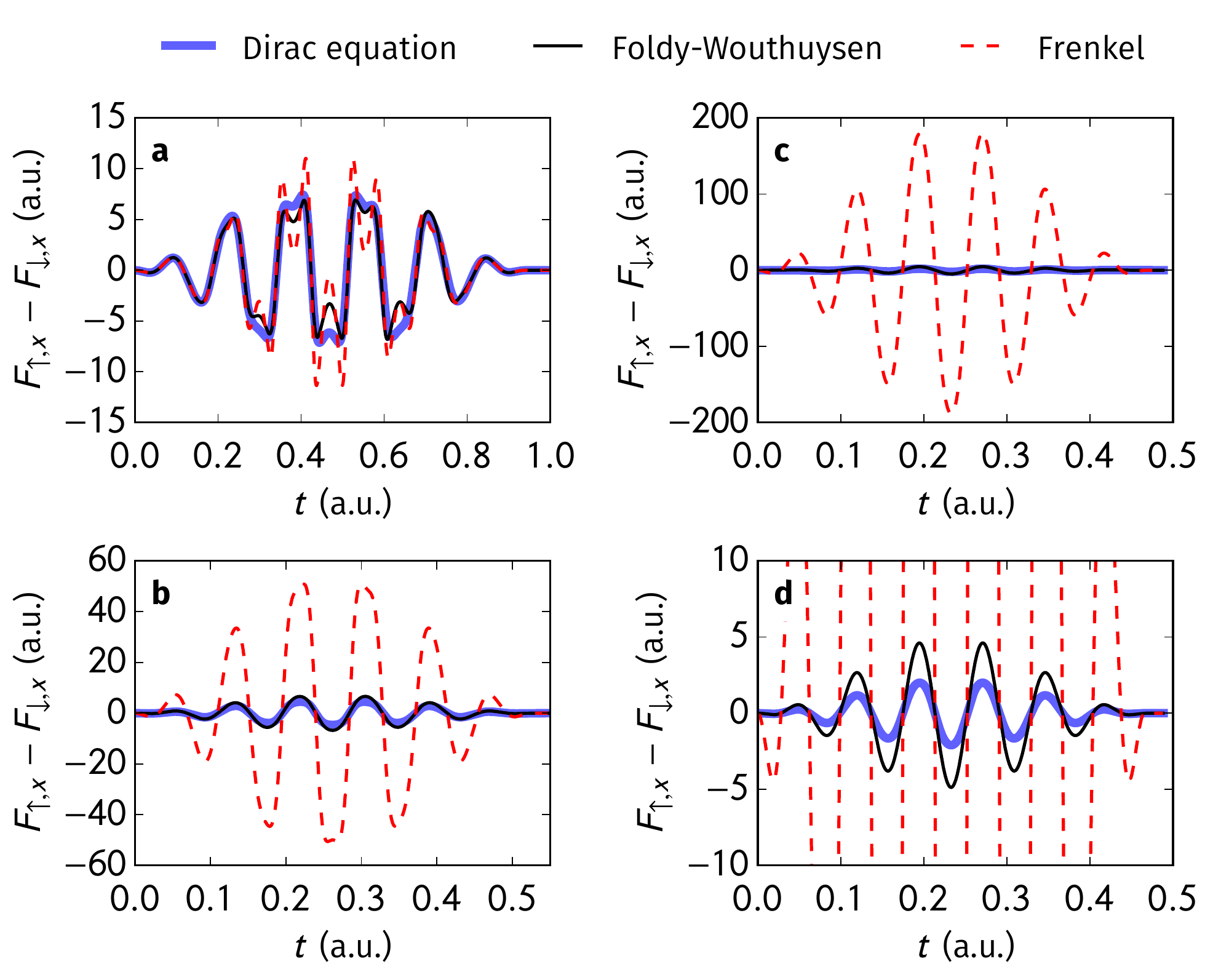}
  \caption{Difference between the force components in laser propagation
    direction for electrons in a strong plane-wave laser pulse with
    initial spin orientation parallel or anti-parallel to the magnetic
    field direction as predicted by the various considered models.  The
    four sub-figures correspond to different initial electron momenta
    opposite to the propagation direction of the laser pulse. (a):
    $\vec p=\trans{(0,0,0)}$; (b): $\vec p=\trans{(-mc,0,0)}$; (c) and
    (d): $\vec p=\trans{(-2mc,0,0)}$. Sub-figures (c) and (d) show the
    same data but on different scales.  Laser parameters are peak
    electric field strength $\hat E=
    \mbox{$2.57\times10^{15}\,\mathrm{V/m}$}$, wavelengths
    $\lambda=
    1.06\,\mathrm{nm}$,
    the pulse length equals $n=6$ cycles.  In case of the Dirac equation
    the wavepacket had an initial width of $
    0.026\,\mathrm{nm}$.}
  \label{fig:free_partice_laser_force}
\end{figure*}

The dynamics of a classical particle with rest mass $m$, charge $q$, and
internal spin degree of freedom are governed by the modified Lorentz
equation
\begin{subequations}
  \label{eq:dynamics}
  \begin{equation}
    \label{eq:dudt}
    M\frac{\D u^\alpha}{\D \tau} =
    q F^{\alpha\beta} u_\beta + F_\mathrm{s}^\alpha \,,
  \end{equation}
  which incorporates the model-specific spin-dependent force
  $F_\mathrm{s}^\alpha= \gamma
  \trans{(F_\mathrm{s}^0,\vec{F}_\mathrm{s})}$,
  and the Thomas-Bargmann-Michel-Telegdi equation
  \cite{Thomas:1926:Motion_of,Bargmann:Michel:etal:1959:Precession_of}
  \begin{equation}
    \label{eq:TBMT}
    \frac{\D S^\alpha}{\D \tau} = 
    \frac{q}{m} 
    F^{\alpha\beta}S_\beta 
    \,.
  \end{equation}
\end{subequations}
Here, $\tau$ denotes the proper time of the particle with
$\D\tau=\D t/\gamma$,
$u_\alpha=\D r_\alpha/\D\tau=\trans{(\gamma c,-\vec{p}/m)}$ the
four-velocity, $r_\alpha=\trans{(ct,-\vec{r})}$ the time-space
coordinate, $\gamma=\sqrt{1+\vec{p}^2/(mc)^2}$ the relativistic Lorentz
factor, $c$ the speed of light, $F^{\alpha\beta}$ the electrodynamic
field tensor, $\vec{p}$ the kinematic momentum, $M$ the effective mass,
$S^\alpha=\trans{(S^0, \vec{S})}$ the spin's four-vector in the
laboratory frame with
\begin{align}
  S^0 & = \dfrac{\vec{p}\cdot\vec{s}}{mc} 
  && \text{and} &
  \vec{S} & = 
  \vec{s} + \dfrac{\vec{p}\cdot\vec{s}}{(\gamma+1)m^2c^2}\vec{p} \,.
\end{align}
The classical spin vector in the rest frame $\vec s$ of length $\hbar/2$
is proportional to the particle's polarisation and corresponds to the
spin operator in quantum mechanics.  The spin-dependent forces may be
written for the classical Foldy-Wouthuysen model (FW) and the Frenkel
model (F) as
\cite{Silenko:2008:Foldy-Wouthyusen_transformation,Frenkel:1926:Elektrodynamik_des}
\begin{subequations}
  \begin{align}
  \label{coeq:F_FW}
    F_\mathrm{s,\,FW}^\alpha & =
    - \frac qm \frac{\partial}{\partial r_\alpha} U_\mathrm{FW}\,,\\
    \label{coeq:F_F}
    F_\mathrm{s,\,F}^\alpha & = 
    -\frac qm \left(\frac{\partial}{\partial r_\alpha} -
    \frac{u^\alpha u^\beta}{c^2} \frac{\partial}{\partial r^\beta}\right) U_\mathrm{F}\,,
  \end{align}
\end{subequations}
with the scalars $U_\mathrm{FW}$ and $U_\mathrm{F}$ defined as
\begin{subequations}
  \begin{align}
  U_\mathrm{FW} & =\vec{s} \cdot \left(
    \vec{B} -
    \frac{\vec{p}\times\vec{E}}{(\gamma+1)mc^2}
  \right) \,, \\
   U_\mathrm{F} & =\vec{s} \cdot \left(
     \gamma\vec B - \frac{1}{mc^2}\,\vec{p}\times\vec{E} + 
     \frac{\vec{p}\cdot\vec{B}}{(\gamma+1)m^2c^2}\vec{p}
   \right)\,.
  \end{align}
\end{subequations}
The effective masses in these two models are $M_\mathrm{FW}=m$ and
\mbox{$M_\mathrm{F}= m-q\gamma U_\mathrm{F}/(mc^2)$}, respectively.  The
forces \eqref{coeq:F_FW} and \eqref{coeq:F_F} become equal in the limit
of low electron energies.  The classical Foldy-Wouthuysen and the
Frenkel models differ mainly in the large-kinematic-momentum limit.

The spin-dependent forces \eqref{coeq:F_FW} and \eqref{coeq:F_F} are
gradient forces, which become large in systems of ultra strong laser
fields in the short-wavelength limit.  In the following, we consider the
interaction of relativistic electrons with strong electromagnetic fields
in the X-ray regime.  The electron moves initially in $x$~direction
opposite to the plane-wave laser pulse, which is assumed to have linear
polarisation in $y$~direction and to be modulated by a $\sin^2$-shaped
envelope.  At time zero, the front of the laser pulse reaches the origin
of the coordinate system, where the electron is initially located.  The
electron's initial spin orientation is parallel or anti-parallel to the
direction of the magnetic field ($z$ direction), representing spin up
(indicated by $\uparrow$) or down (indicated by $\downarrow$) states.
Note that as a consequence of equation~\eqref{eq:TBMT} the spin remains
in its initial state for all times for the considered setup.

Due to the spin-dependent forces, the electron's trajectory depends on
the spin orientation.  Although the influence of the spin on the shape
of the trajectory is very small, the spin-dependent force may be used to
compare the three models.  In the classical models the total force can
be split directly into a Lorentz force part and a spin force part (see
equation~\eqref{eq:dudt}), which is not possible in the framework of the
Dirac equation.  Therefore, the magnitude of the spin-dependent force is
evaluated by calculating the difference of the total forces
$\vec F_{\uparrow}$ and $\vec F_{\downarrow}$ for the trajectories of
electrons with initial spin parallel and anti-parallel to the
$z$~direction.  Figure~\ref{fig:free_partice_laser_force} shows the
$x$~component of \mbox{$\vec F_{\uparrow}-\vec F_{\downarrow}$} as a
function of time $t$ as determined from the Dirac equation, the
classical Foldy-Wouthuysen model, and the Frenkel model.  For zero
initial momentum, the three models yield very similar results, as shown
in figure~\ref{fig:free_partice_laser_force}(a).  In particular, the
predictions of the Dirac equation and the classical Foldy-Wouthuysen
model for the force difference \mbox{$F_{\uparrow,x}-F_{\downarrow,x}$}
match very well and the prediction of the Frenkel model shows only small
deviations from the other two models.  The qualitative predictions of
the various models diverge with growing initial electron momentum, see
\mbox{figure~\ref{fig:free_partice_laser_force}(b)--(d)}.  The Frenkel
model predicts that the force difference becomes larger for relativistic
electrons, while the Dirac equation and the classical Foldy-Wouthuysen
model yield smaller force differences.  The qualitatively different
behaviour of the Frenkel model and the classical Foldy-Wouthuysen model
is a consequence of a different dependence on $\gamma$ of the
spin-dependent forces~\eqref{coeq:F_FW} and~\eqref{coeq:F_F}.

For symmetry reasons the net effect of the plane-wave pulse on the
electron momentum vanishes in the Frenkel model as well as the classical
Foldy-Wouthuysen model, although both models predict different forces
acting on the electron during its interaction with the laser pulse.
Therefore, a plane-wave setup is not suitable to test the classical
models experimentally.  However, considering focused infrared laser
pulses of upcoming high-power laser facilities the discrepancy in the
predicted electron dynamics by the classical Foldy-Wouthuysen and the
Frenkel models becomes large enough to distinguish between them
experimentally.  

An electron, which is initially directed towards the focus of a
counter-propagating high-intensity laser pulse with linear polarisation,
is displaced transversely due to the transverse electric field.  When
the oscillating field changes its sign, the force drives the electron
back to its initial transverse position.  However, this force is smaller
than the expelling force due to the focusing inhomogeneity.  As a
result, the oscillation centre of a spinless charged particle drifts
radially from the spot centre, which is called ponderomotive scattering
\cite{Bucksbaum:Bashkansky:etal:1987:Scattering_of}.  Beside the
deflection of a charged particle in the ponderomotive potential of the
laser fields, the spin may induce a further deflection via the
spin-dependent forces \eqref{coeq:F_FW} and \eqref{coeq:F_F}, in
particular, if the electron is polarised parallel or anti-parallel to
the direction of the magnetic field.  As the spin is
\mbox{(anti-)parallel} to the magnetic field direction it follows from
equation~\eqref{eq:dynamics} that the electron remains in the plane
perpendicular to the magnetic field direction and the electron's spin is
frozen to its initial state.  The deflection of a particle in the
ponderomotive potential of the focused laser pulse is defined by the
angle $\theta$ between its initial momentum and its final momentum after
the particle is separated from the laser fields.  It is dominated by the
ponderomotive scattering due to the dominant Lorentz force, which
increases with increasing field strength.

\begin{figure}
  \centering
  \includegraphics[scale=0.475]{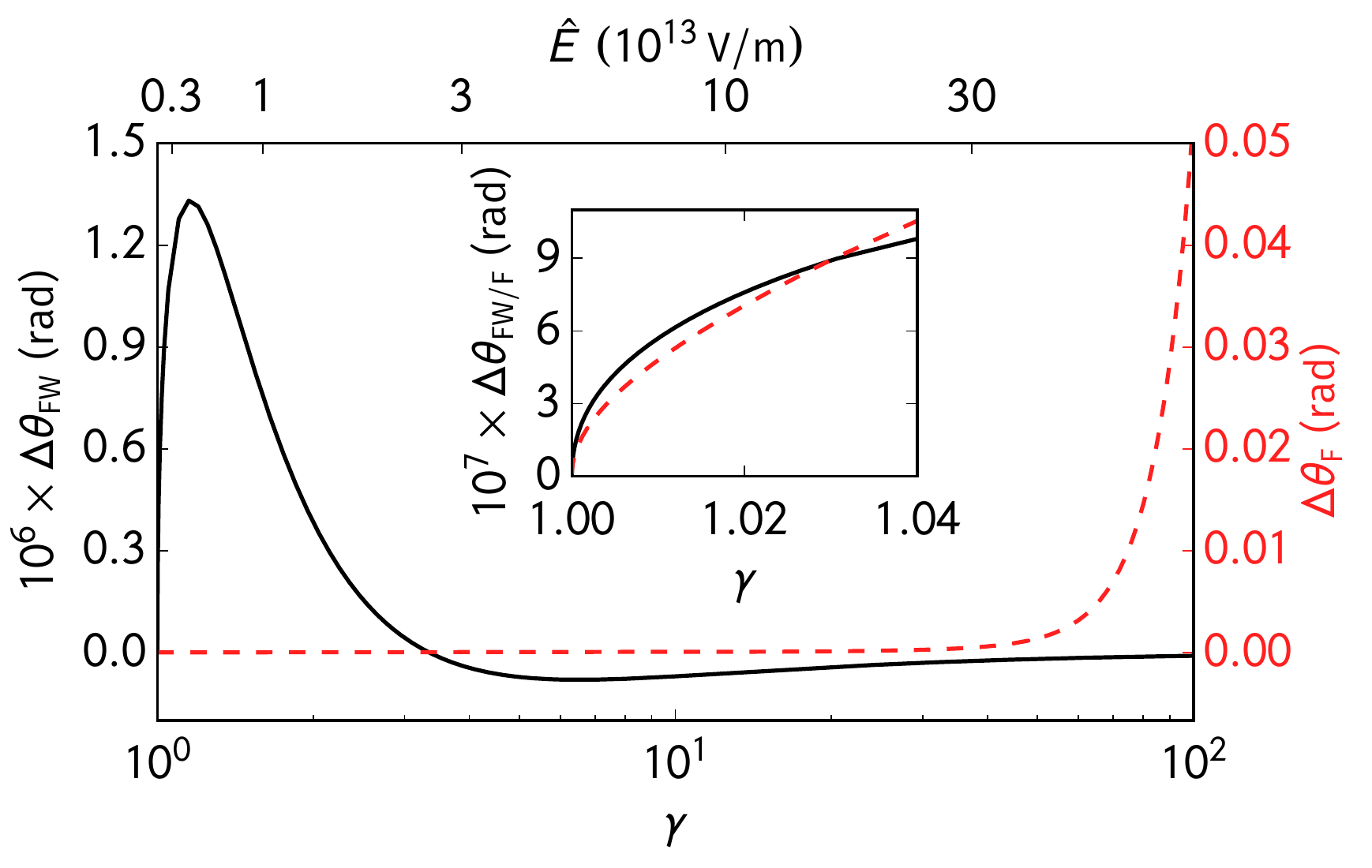}
  \caption{The aberration angle $\Delta\theta$ between spin-up and
    spin-down electrons induced by the ponderomotive potential as a
    function of the initial energy $\gamma mc^2$ of the particle for the
    classical Foldy-Wouthuysen (solid black line, left scale) and
    Frenkel (dashed light red line, right scale) models. The inset shows
    the non-relativistic limit.  The electric field strength of the
    counter-propagating laser pulse scales with initial $\gamma$ as
    $\hat E=4\pi mc^2\sqrt{\gamma^2-1}/(|q|\lambda)$, which causes a
    strong acceleration of the electron opposite to its initial velocity
    but without reflecting it.  Other parameters are the wavelength
    $\lambda=
    800\,\mathrm{nm}$,
    the duration (number of cycles) $n=20$, and focus radius
    $w_0=2\lambda$.}
  \label{fig:focussed_pulse_theta}
\end{figure}

Besides the deflection due to the charge, the electron's spin state
leads to a modification of the deflection angle $\theta$, which depends
on the spin orientation.  In this way, one can define the aberration
angle $\Delta \theta = \theta_\uparrow - \theta_\downarrow$, where
$\theta_\uparrow$ and $\theta_\downarrow$ denote the deflection angles
for the spin-up and spin-down cases, see also Methods section.  The
Frenkel and the classical Foldy-Wouthuysen models lead to different
aberrations $\Delta \theta$, as shown in
figure~\ref{fig:focussed_pulse_theta} for varying electron energies.  As
indicated in the inset, the two models share the same non-relativistic
limit.  In the relativistic regime, the angle $\Delta \theta$ as
predicted by the classical Foldy-Wouthuysen model does not vary with the
electron's initial energy monotonically and it may even change its
sign.  Furthermore, the absolute value of the spin-induced additional
deflection angle $\Delta\theta_\mathrm{FW}$ from the classical
Foldy-Wouthuysen model remains under the magnitude of
$10^{-6}\,\mathrm{rad}$ and decreases with the electron's initial energy
in the relativistic parameter regime.  In contrast to the classical
Foldy-Wouthuysen model, the aberration angle of the Frenkel model
$\Delta \theta_\mathrm{F}$ increases to about $0.05\,\mathrm{rad}$ with
the electron's energy for relativistic electrons in high-intensity laser
fields of the applied parameters.

\section*{Discussion}

We have investigated the dynamics of electrons in various setups by
applying two different classical models, the classical Foldy-Wouthuysen
and the Frenkel models.  The predictions of these classical models were
compared to each other and to predictions by the Dirac equation, when a
numerical solution of the Dirac equation was feasible.  In specific
parameter regimes, these classical models can lead to conflicting
predictions.  The Frenkel model may be of timely interest
\cite{Deriglazov:2014:Lagrangian_for,Mahajan:Asenjo:etal:2014:Comparison_of,Flood:Burton:2015:Stern-Gerlach_surfing}
and prominent \cite{Barut:1980:Electrodynamics,Corben:1968} for its much
longer history and its wide application
\cite{Jackson:2008:Examples_of,Hushwater:2014:On_discovery}.  The
classical Foldy-Wouthuysen model, however, may be superior as it is
qualitatively in better agreement with the quantum mechanical Dirac
equation.

The discrepancies in the predictions of the two classical models may
become experimentally detectable in light-matter interaction in strong
highly focused beams.  As electron bunches with the emittance as low as
$10^{-3}\,\mathrm{rad}$ have been prepared
\cite{PhysRevLett.109.064802}, the spin-induced aberration angle of the
order of $10^{-2}\,\mathrm{rad}$ from the Frenkel model is potentially
measurable, if an electron beam with an energy of tens of MeV and an
infrared laser of the intensity $\sim 10^{22}\,\mathrm{W/cm^2}$ are
applied as discussed above.  The spin-induced contribution to the
deflection as predicted by the classical Foldy-Wouthuysen model, which
is for the applied parameters of the order of $10^{-6}\,\mathrm{rad}$,
is too small to be demonstrated.  However, a differentiation among both
predictions appears feasible.  In current head-on experiments
\cite{Burke:Field:etal:1997:Positron_Production,Ta-Phuoc:Corde:etal:2012:All-optical_Compton}
with focused fields of high inhomogeneities and energetic electrons no
significant spin effect in orbital motion was observed.  The lack of
experimental evidence for a non-negligible spin-induced deflection may
be seen as a superiority of the classical Foldy-Wouthuysen model again
regarding spin modified dynamics.

\section*{Methods}

For the plane-wave setup, the laser pulse is assumed to have linear
polarisation in the $y$ direction and is modulated by a $\sin^2$-shaped
envelope
\begin{equation}
  \label{eq:sin2_env}
  w(\eta) = \theta(-\eta)\theta(\eta+\pi)\sin^2\eta \,,
\end{equation}
where $\theta(\eta)$ denotes the Heaviside step function.  Introducing
the wavelength $\lambda$, the peak amplitude $\hat E$, and the pulse
width $n$ measured in laser cycles, the electric field component of
the laser pulse is given by
\begin{align}
  \label{eq:E_pulse}
  \vec E(\vec r, t) & = 
  \hat{E}\sin\frac{2 \pi(x-ct)}\lambda \,
  w\left(\frac{\pi(x-ct)}{n\lambda }\right) \vec e_y 
\end{align}
and the magnetic field component follows via
$\vec B(\vec r, t)=\mbox{$\vec\e_x\times \vec E(\vec r, t)/c$}$.  At
time zero the front of the laser pulse reaches the origin of the
coordinate system, where the electron is initially located.  The
electron's initial spin orientation is parallel or anti-parallel to the
direction of the magnetic field ($z$~direction).  We solved the
equations of motion of the classical Foldy-Wouthuysen model and the
Frenkel model for the plane-wave setup numerically via the Boris
algorithm \cite{Birdsall:Langdon:2004:Plasma_Physics}.  The
time-dependent Dirac equation for a two-dimensional wavepacket in the
same setup was propagated numerically employing a Fourier split operator
method
\cite{Braun:Su:etal:1999:Numerical_approach,Bauke:Keitel:2011:Accelerating_Fourier,Fillion-Gourdeau:Lorin:etal:2012:Numerical_solution,Fillion-Gourdeau:Lorin:etal:2014:split-step_numerical}.
In order not to violate the quantum-classical correspondence between
classical operators and quantum mechanical operators, the Dirac
wavepacket was prepared to have a small width compared to the wavelength of
the applied electromagnetic field
\cite{Chen:Chiou:2014:High-order_Foldy-Wouthuysen}.  The force, which
acts on the electron during its interaction with the plane-wave
electromagnetic field and which enters in
figure~\ref{fig:free_partice_laser_force}, is given by
equations~\eqref{coeq:F_FW} and~\eqref{coeq:F_F}.  In the case of the
Dirac equation, the force was determined as the time derivative of the
quantum mechanical expectation value of the electron's kinematic
momentum
$\braket{\Psi(\vec r, t)|-\I\hbar\vec\nabla - q \vec A(\vec r,
  t)|\Psi(\vec r, t)}$,
where $\Psi(\vec r, t)$ is the electron's four-component wave function
and $\vec A(\vec r, t)$ the vector potential of the electromagnetic
fields.

For the setup with a focused infrared laser pulse, numerical solutions
of the Dirac equation are not feasible due to the long time scale of
infrared laser pulses.  A longer pulse length in combination with
wavepacket spreading leads to a $\lambda^3$- or $\lambda^4$-scaling of
the computational demand to solve the Dirac equation in two or
respectively three dimensions.  Thus numerical simulations were limited
to the two classical models.

\begin{figure}
  \centering
  \includegraphics[scale=0.45]{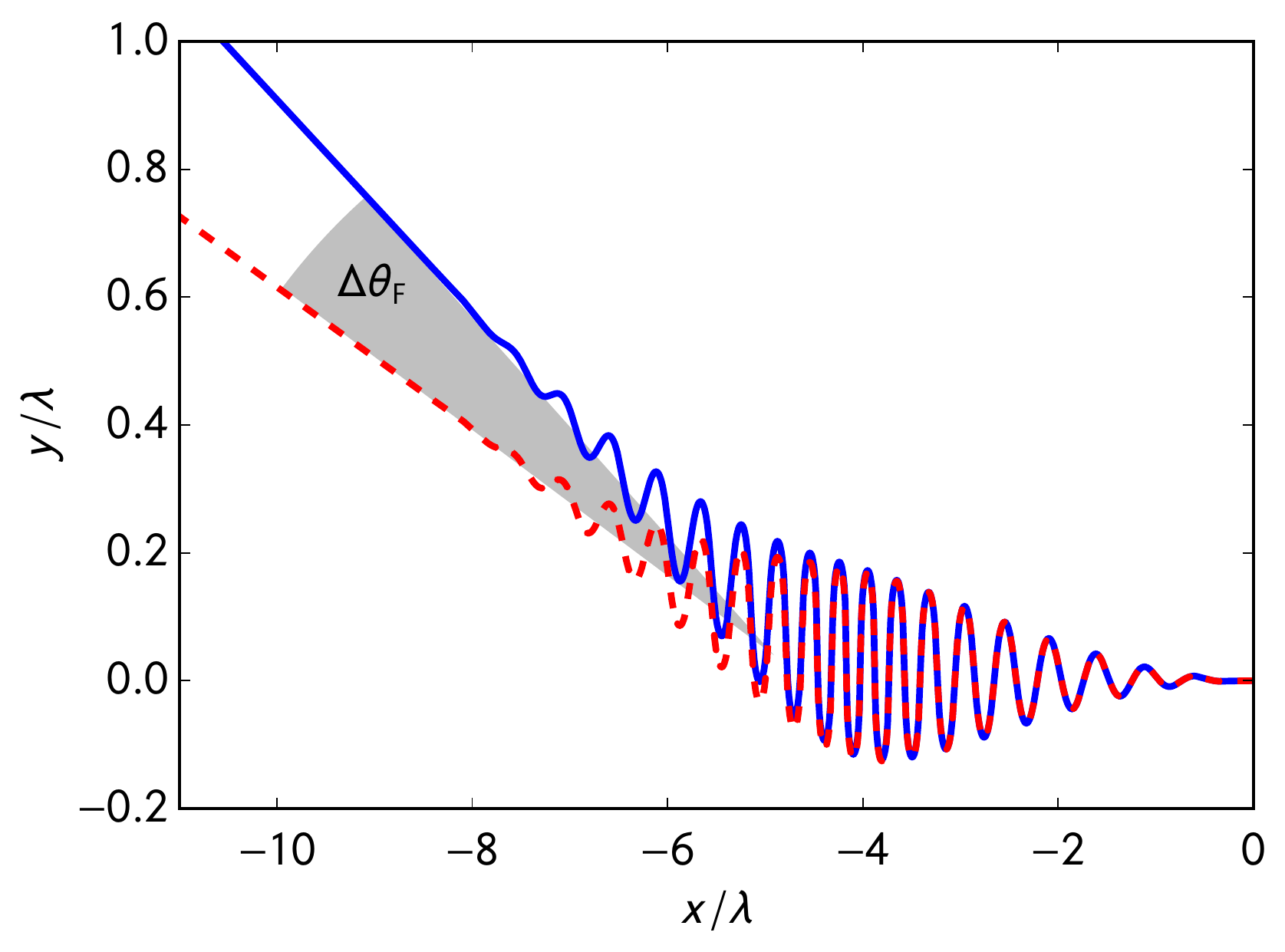}
  \caption{Definition of the aberration angle $\Delta\theta$.  Lines
    represent trajectories of a highly energetic electron with initial
    $\gamma=100$ in a focused laser pulse with the wavelength
    $\lambda=800\,\mathrm{nm}$, the amplitude of strength
    $\hat E = 8.03\times10^{14}\,\mathrm{V/m}$, the duration (number of
    cycles) $n=20$, and focus radius $w_0=2\lambda$.  The solid and
    dashed curves correspond to those of electrons with the spin
    parallel to the $z$ axis and anti-parallel to the $z$ axis (magnetic
    field direction), respectively, as described by the Frenkel model.
    The deflection angle for the Frenkel model is indicated by
    $\Delta\theta_\mathrm{F}$.}
  \label{fig:focussed_pulse_trajectory}
\end{figure}

The polarisation and the longitudinal (in propagation direction) profile
of the focused laser pulse are as in the plane-wave case.  The
transverse profile and the phase are modelled as a Gaussian beam with the
transversal focus radius $w_0$, e.\,g.\ with terms up to the 5th order
of the small diffraction angle $\epsilon=w_0/x_r$ as in
Ref.~\cite{Salamin:Keitel:2002:Electron_Acceleration}, where
$x_r=\pi w_0^2/\lambda$ is the Rayleigh length.  The phase of the
focused pulse depends not only on the longitudinal coordinate but also
on the transverse coordinate.  The deflection of an electron in a
head-on collision with a focused laser pulse is defined by the angle
between the final transverse and the longitudinal momentum components
after the particle is separated from the laser fields.  Due to
spin-dependent forces the deflection depends on the electron's initial
spin orientation (relative to the magnetic field direction).  In this
way, the aberration angle $\Delta \theta$ is defined as the angle
between the final momenta for electrons with initial spin-up and
spin-down orientation, see figure~\ref{fig:focussed_pulse_trajectory}.

The considered setup with the employed parameters is also sensitive to
radiative reaction forces.  Our calculations involving both spin and
radiative reaction forces (via the Landau-Lifshitz equation
\cite{Landau:Lifshitz:II}) have, however, confirmed that the key
deviations displayed in figure~\ref{fig:focussed_pulse_theta} are not
essentially modified such as especially the strong rise of the
aberration angle $\Delta\theta$ for the Frenkel model as compared to the
Foldy-Wouthuysen model in the highly relativistic regime.  Because of
this and since the Landau-Lifshitz equation has not been confirmed
experimentally as well, we decided here to present comparisons not
including the radiative reaction forces.

\bibliography{Frenkel}

\section*{Competing financial interests}

The authors declare no competing financial interests.

\section*{Author contributions}

H.~B. performed the numerical solution of the Dirac equation and
implemented the necessary numerical methods.  Numerical simulations for
the Frenkel model and the classical Foldy-Wouthuysen model have been
carried out by M.~W.  All authors contributed to the design of the
project, interpretation of the data and the writing of the paper.

\end{document}